\documentclass[twocolumn,superscriptaddress,aps,longbibliography,showpacs,preprintnumbers,amsmath,amssymb,floatfix]{revtex4-1}
\usepackage{titlesec}
\usepackage{amsmath}
\usepackage{graphicx}
\usepackage[ansinew]{inputenc}
\usepackage{array}
\usepackage{color}
\usepackage{amsxtra}
\usepackage{amstext}
\usepackage{amssymb}
\usepackage{latexsym}
\usepackage{gensymb}
\usepackage{dsfont}
\usepackage{braket}
\usepackage[caption=false]{subfig}
\usepackage[makeroom]{cancel}

\usepackage{dcolumn}
\usepackage{bm}
\usepackage{bbm}
\usepackage{braket}
\usepackage{mathrsfs}
\usepackage{amstext}
\usepackage{euscript}
\usepackage{txfonts}
\usepackage{soul}

\usepackage[unicode=true,
 bookmarks=false,
 breaklinks=false,pdfborder={0 0 1},backref=false,colorlinks=true,citecolor=blue]
 {hyperref}

\newcommand{\ms}[1]{\mbox{\scriptsize #1}}

\makeatletter
\@ifundefined{textcolor}{}
{%
 \definecolor{BLACK}{gray}{0}
 \definecolor{WHITE}{gray}{1}
 \definecolor{RED}{rgb}{1,0,0}
 \definecolor{GREEN}{rgb}{0,1,0}
 \definecolor{BLUE}{rgb}{0,0,1}
 \definecolor{CYAN}{cmyk}{1,0,0,0}
 \definecolor{MAGENTA}{cmyk}{0,1,0,0}
 \definecolor{YELLOW}{cmyk}{0,0,1,0}
}

\makeatletter
\renewcommand*\env@matrix[1][*\c@MaxMatrixCols c]{%
  \hskip -\arraycolsep
  \let\@ifnextchar\new@ifnextchar
  \array{#1}}
\makeatother

\newcommand{\cref}[1]{Ref.\,\cite{#1}}

\makeatother
\RequirePackage[normalem]{ulem} 
\RequirePackage{color}\definecolor{RED}{rgb}{1,0,0}\definecolor{BLUE}{rgb}{0,0,1} 
\providecommand{\DIFaddbegin}{} 
\providecommand{\DIFaddend}{} 
\providecommand{\DIFdelbegin}{} 
\providecommand{\DIFdelend}{} 
\providecommand{\DIFaddbeginFL}{} 
\providecommand{\DIFaddendFL}{} 
\providecommand{\DIFdelbeginFL}{} 
\providecommand{\DIFdelendFL}{} 
\newcommand{\DIFscaledelfig}{0.5}
\RequirePackage{settobox} 
\RequirePackage{letltxmacro} 
\newsavebox{\DIFdelgraphicsbox} 
\newlength{\DIFdelgraphicswidth} 
\newlength{\DIFdelgraphicsheight} 
\LetLtxMacro{\DIFOincludegraphics}{\includegraphics} 
\newcommand{\DIFaddincludegraphics}[2][]{{\color{blue}\fbox{\DIFOincludegraphics[#1]{#2}}}} 
\newcommand{\DIFdelincludegraphics}[2][]{
\sbox{\DIFdelgraphicsbox}{\DIFOincludegraphics[#1]{#2}}
\settoboxwidth{\DIFdelgraphicswidth}{\DIFdelgraphicsbox} 
\settoboxtotalheight{\DIFdelgraphicsheight}{\DIFdelgraphicsbox} 
\scalebox{\DIFscaledelfig}{
\parbox[b]{\DIFdelgraphicswidth}{\usebox{\DIFdelgraphicsbox}\\[-\baselineskip] \rule{\DIFdelgraphicswidth}{0em}}\llap{\resizebox{\DIFdelgraphicswidth}{\DIFdelgraphicsheight}{
\setlength{\unitlength}{\DIFdelgraphicswidth}
\begin{picture}(1,1)
\thicklines\linethickness{2pt} 
{\color[rgb]{1,0,0}\put(0,0){\framebox(1,1){}}}
{\color[rgb]{1,0,0}\put(0,0){\line( 1,1){1}}}
{\color[rgb]{1,0,0}\put(0,1){\line(1,-1){1}}}
\end{picture}
}\hspace*{3pt}}} 
} 
\LetLtxMacro{\DIFOaddbegin}{\DIFaddbegin} 
\LetLtxMacro{\DIFOaddend}{\DIFaddend} 
\LetLtxMacro{\DIFOdelbegin}{\DIFdelbegin} 
\LetLtxMacro{\DIFOdelend}{\DIFdelend} 
\DeclareRobustCommand{\DIFaddbegin}{\DIFOaddbegin \let\includegraphics\DIFaddincludegraphics} 
\DeclareRobustCommand{\DIFaddend}{\DIFOaddend \let\includegraphics\DIFOincludegraphics} 
\DeclareRobustCommand{\DIFdelbegin}{\DIFOdelbegin \let\includegraphics\DIFdelincludegraphics} 
\DeclareRobustCommand{\DIFdelend}{\DIFOaddend \let\includegraphics\DIFOincludegraphics} 
\LetLtxMacro{\DIFOaddbeginFL}{\DIFaddbeginFL} 
\LetLtxMacro{\DIFOaddendFL}{\DIFaddendFL} 
\LetLtxMacro{\DIFOdelbeginFL}{\DIFdelbeginFL} 
\LetLtxMacro{\DIFOdelendFL}{\DIFdelendFL} 
\DeclareRobustCommand{\DIFaddbeginFL}{\DIFOaddbeginFL \let\includegraphics\DIFaddincludegraphics} 
\DeclareRobustCommand{\DIFaddendFL}{\DIFOaddendFL \let\includegraphics\DIFOincludegraphics} 
\DeclareRobustCommand{\DIFdelbeginFL}{\DIFOdelbeginFL \let\includegraphics\DIFdelincludegraphics} 
\DeclareRobustCommand{\DIFdelendFL}{\DIFOaddendFL \let\includegraphics\DIFOincludegraphics} 

\begin{document}


\title{Operational Resource Theory of Nonclassicality via Quantum Metrology}

\author{Wenchao Ge}
\affiliation{Institute for Quantum Science and Engineering (IQSE) and Department of Physics and Astronomy, Texas A\&M University, College Station, TX 77843-4242, USA}
\affiliation{United States Army Research Laboratory, Adelphi, Maryland 20783, USA}
\affiliation{The Institute for Research in Electronics and Applied Physics (IREAP), College Park, Maryland 20740, USA}

\author{Kurt Jacobs}
\affiliation{United States Army Research Laboratory, Adelphi, Maryland 20783, USA}
\affiliation{Department of Physics, University of Massachusetts at Boston, Boston, Massachusetts 02125, USA}
\affiliation{Hearne Institute for Theoretical Physics, Louisiana State University, Baton Rouge, Louisiana 70803, USA}

\author{Saeed Asiri}
\affiliation{The National Center for Laser and Optoelectronics Technologies, KACST, Riyadh 11442, Saudi Arabia}
\affiliation{Center for Quantum Optics and Quantum Informatics (CQOQI), KACST, Riyadh 11442, Saudi Arabia}

\author{Michael Foss-Feig}
\affiliation{United States Army Research Laboratory, Adelphi, Maryland 20783, USA}
\affiliation{Joint Quantum Institute, NIST/University of Maryland, College Park, Maryland 20742, USA}
\affiliation{Joint Center for Quantum Information and Computer Science, NIST/University of Maryland, College Park, Maryland 20742, USA}

\author{M. Suhail Zubairy}
\affiliation{Institute for Quantum Science and Engineering (IQSE) and Department of Physics and Astronomy, Texas A\&M University, College Station, TX 77843-4242, USA}

\date{\today}

 \begin{abstract}
 The nonclassical properties of quantum states are of tremendous interest due to their potential applications in future technologies. It has recently been realized that the concept of a ``resource theory" is a powerful approach to quantifying and understanding nonclassicality. To realize the potential of this approach one must first find resource theoretic measures of nonclassicality that are ``operational", meaning that they also quantify the ability of quantum states to provide enhanced performance for specific tasks, such as precision sensing. Here we achieve a significant milestone in this endeavor by presenting the first such operational resource theoretic measure. In addition to satisfying the requirements of a resource measure, it has the closest possible relationship to the quantum-enhancement provided by a non-classical state for measuring phase-space displacement: it is equal to this enhancement for pure states, and is a tight upper bound on it for mixed states. We also show that a lower-bound on this measure can be obtained experimentally using a simple Mach-Zehnder interferometer.
 \end{abstract} 
 \maketitle


For a single optical mode, or equivalently a linear oscillator, it is the coherent states that are regarded as the quantum equivalent of classical states, due to the fact that they have minimal uncertainty in phase and amplitude, and apart from the uncertainty principle behave classically. The ability of quantum systems to exist in a superposition of two or more distinguishable states is arguably their most intriguing feature~\cite{Raimond01}. A superposition of coherent states, which may have highly nonclassical features, can enable quantum-enhanced technologies in communications~\cite{Braunstein98}, computation~\cite{Braunstein05, Knill:2001aa}, metrology \cite{giovannetti2006quantum}, and is useful in probing the limits of quantum mechanics~\cite{Leggett:2002aa, Arndt:2014aa}. Moreover, single-mode nonclassical states can be converted into multi-mode entangled states using linear optics \cite{Kim:02, Wang:02, Vogel:14, BrunelliPRA15, ge2015conservation,Killoran:2016aa}. Given the usefulness of ``nonclassicality", finding ways to quantify it may provide further insights into its features and the resources required to create and manipulate it.  

A single-mode quantum state $\hat{\rho}$ can be represented by the Glauber-Sudarshan $P$ function \cite{Glauber63, Sudarshan63} as 
\begin{eqnarray}
\hat{\rho}=\int P(\alpha,\alpha^{\ast})\ket{\alpha}\bra{\alpha}d^2\alpha.
\label{eq:P-fucntion}
\end{eqnarray}
The state $\hat\rho$ is defined as classical if the distribution function $P(\alpha,\alpha^{\ast})$ is positive-definite, in which case it merely represents a classical probability density over the coherent states $|\alpha\rangle$. Otherwise the state is said to be nonclassical. 

Since the $P$ function itself does not provide a quantitative measure of nonclassicality~\cite{tan2019negativity}, a number of approaches have been taken to obtaining one~\cite{hillery1987nonclassical, Marian:02,Lee:91,AsbothPRL05,GehrkePRA12, Vogel:14, Arkhipov:2016aa,YuanPRA18}. Measures proposed include the ``nonclassical distance"~\cite{hillery1987nonclassical, Marian:02,Nair17}, the ``nonclassical depth" \cite{Lee:91}, the ``entanglement potential" \cite{AsbothPRL05}, and the Schmidt rank \cite{GehrkePRA12, Vogel:14, Ryl17}. 

It was realized recently that the notion of nonclassicality can be put on a much surer footing by using a resource theory~\cite{Tan17, Baumgratz14, coecke16, Streltsov17,RevModPhys.91.025001}. Resource theories define the quantity of interest by specifying the operations under which it should be impossible to increase it. These are referred to as the ``free" operations. For entanglement it is local operations and classical communication that are free, while for nonclassicality it is the class of operations that does not allow one to create superpositions of coherent states from mixtures of them (these operations will be discussed further below). To obtain a resource theory of nonclassicality one has ``merely" to find a measure of nonclassicality that does not increase under the free operations.

While resource theory alone provides a well-motivated foundation, more important is a resource theory that also quantifies the ability of a state to perform useful tasks (metrology, cryptography, etc.), because it can provide insight into the non-classical resources underlying the quantum-enhancement of these tasks. We will refer to a resource theoretic measure that does so as an \textit{operational resource theory}. Recently some progress has been made in this direction by two works considering the quantum-enhanced measurement of quadrature (equivalently, the measurement of displacement in phase space)~\cite{YadinPRX18,Kwon19}. However, neither of these works were able to achieve an operational resource measure. Of the potential measures investigated, one was a bona-fide resource measure but did not capture the quantum-enhanced precision for the mixed states~\cite{Kwon19}, and the other captured the precision~\cite{YadinPRX18} but did not satisfy the minimal requirements of a resource measure~\cite{Streltsov17}.

Here we present the first operational resource measure of nonclassicality. It satisfies the minimal requirements of a resource theory, quantifies the ability to perform quadrature measurement for pure states, and is a tight upper bound for the latter for mixed states. As we discuss below, for mixed states a tight upper bound is the strongest relationship that can be obtained. We also show that a lower bound on our measure can be obtained experimentally using the simplest interferometric configuration, the Mach-Zehnder interferometer (MZI)~\cite{Caves81,Pezze08, TanPRA14, Demkowicz-Dobrzanski:2015aa}. We do this by showing that the MZI, when used in the ``balanced'' configuration, converts the ability of a state to measure quadrature directly into the ability to measure a phase shift. The phase precision of the MZI thus reveals the quadrature precision provided by the input state. 

Nonclassicality has a close relationship with the concept of ``macroscopicity" which attempts to quantify the ``size" of a superposition~\cite{Frowis2018}. The two concepts are very close, because a superposition of macroscopically distinguishable states is a nonclassical feature, and the metrological power of a pure quantum state is limited by its energy. As a result, measures of nonclassicality can often serve as good measures of macroscopicity \cite{YadinPRX18,Kwon19}. In the final part of this Letter we discuss our measure of nonclassicality as a measure of macroscopicity.  


\textit{Metrological power of quantum states.---}
To measure a classical parameter, $\theta$, one applies a transformation $e^{i\theta\hat{G}}$ to a system and then measures the system to determine the parameter from the change that the transformation has induced. The precision with which one can measure the parameter depends on the initial state of the system, $\hat\rho$, and is captured by the quantum Fisher information (QFI)~\cite{Braunstein:1994aa, toth2014quantum, Demkowicz-Dobrzanski:2015aa}. For a pure state, $|\psi\rangle$, the QFI for a parameter $\theta$ and transformation $U(\theta) = e^{i\theta\hat{G}}$ is four times the variance of $\hat{G}$ in state $|\psi\rangle$. For convenience here we will drop the factor of four and define the QFI, denoted by $F_{\hat{G}}(|\psi\rangle)$, simply as the variance: $F_{\hat{G}}(|\psi\rangle) = \langle\psi|(\Delta \hat{G})^2|\psi\rangle$ where $\Delta \hat{G} \equiv \hat{G} - \langle \hat{G} \rangle$. For mixed states the QFI is the convex roof of the variance~\cite{toth2014quantum, Demkowicz-Dobrzanski:2015aa}, namely
\begin{align} 
F_{\hat{G}}\left(\hat{\rho}\right) & = \min_{\{p_j,\ket{\psi_j}\}}\biggl\{\sum_j p_j\langle \psi_j | (\Delta\hat{G})^2 | \psi_j \rangle \biggr\} , 
\end{align} 
in which the minimization is over all ensembles $\{p_j,\ket{\psi_j}\}$,  for which $\hat\rho = \sum_j p_j\ket{\psi_j}\bra{\psi_j}$ and $p_j>0$. Here we will be concerned mainly with the QFI for the quadratures, defined by $\hat{X}_{\mu}=i\left(e^{-i\mu}\hat{a}^{\dagger}-e^{i\mu}\hat{a}\right)/\sqrt{2}$ in which $\hat{a}$ is the annihilation operator for the mode and $\mu\in [0,2\pi]$. We will mainly be interested in the QFI for the quadrature for which the QFI is maximal. We will write this simply as $F_X(\hat\rho)$, which is given by 
\begin{align} 
\label{QFI}
F_X\left(\hat{\rho}\right) & = \max_\mu \Biggl[ \min_{\{p_j,\ket{\psi_j}\}}\biggl\{\sum_j p_j\langle \psi_j | (\Delta\hat{X}_\mu)^2 | \psi_j \rangle \biggr\} \Biggr] . 
\end{align}
We will define the \textit{metrological power} of a quantum state as the amount by which its QFI is greater than the maximum value for any classical state. Since the maximum classical value for the quadrature variance is $1/2$, the metrological power for quadrature measurement is $\mathcal{W}\left(\hat{\rho}\right) \equiv \max[F_X\left(\hat{\rho}\right)-1/2,0]$ and $\mathcal{W}>0$ provides a nonclassicality criterion~\cite{Rivas10}. This operational quantity captures the ability of nonclassical states to enhance metrology, but it is not a measure of nonclassicality because there are mixed non-classical states for which $F_X\left(\hat{\rho}\right) \leq 1/2$.  An example is given by the class of states $\hat{\rho}(p)=\left(1-p\right)\ket{0}\bra{0}+p\ket{1}\bra{1}$ for which $\mathcal{W} = \max\{p(2p-1),0\}$ by calculating $F_X\left(\hat{\rho}\right)$ via the eigenstate decomposition~\cite{toth2014quantum, Demkowicz-Dobrzanski:2015aa}. Because of this property, for mixed states the closest possible  relationship between a resource theoretic measure and the metrological power is that the former provides a tight upper bound on the latter. We note that single-mode Gaussian states are a special class for which $\mathcal{W}$ is a necessary and sufficient condition for nonclassicality~\cite{Kwon19}. 


\textit{Resource theory of nonclassicality.---}
To obtain a resource theory of nonclassicality~\cite{Streltsov17,Tan17}, one must find a measure of nonclassicality, $\mathcal{N}\left(\hat{\rho}\right)$, that satisfies two conditions: (i) Non-negativity: $\mathcal{N}\left(\hat{\rho}\right)\ge0$ for any state $\hat{\rho}$ where the equality holds if and only if $\hat{\rho}$ is classical; (ii) Weak monotonicity: $\mathcal{N}$ cannot be increased by any classical operation $\Lambda$, i.e., $\mathcal{N}\left(\Lambda[\hat{\rho}]\right) \le \mathcal{N}\left(\hat{\rho}\right)$. A classical operation $\Lambda$ is defined as augmentation by any number of classical states (those defined by Eq.\eqref{eq:P-fucntion}), the application of any passive linear optical operations and displacements, and tracing out of the auxiliary modes. The classical operations are the free operations within the resource theory~\cite{Tan17, Kwon19}. 

It has been suggested that a resource theoretic measure of nonclassicality should also satisfy two further conditions~\cite{Tan17, Streltsov17}. These are (iii) strong monotonicity and (iv) convexity. A measure is strongly monotonic when it does not increase, on average, even when a subset of the modes are measured, discarded, and further classical operations are performed dependent on the results of the measurements. Further rounds of measuring and discarding modes and subsequent conditiional operations may also be made. A measure, denoted by $\mathcal{N}\left(\hat{\rho}\right)$, is convex when it obeys $\sum_jp_j\mathcal{N}\left(\hat{\rho}_j\right)\ge\mathcal{N}\left(\sum_jp_j\hat{\rho}_j\right)$ for any quantum states $\hat{\rho}_j$ and probabilities $p_j$.  

To-date, the only proposed metrological measure that is a resource measure (i.e., that has been shown to satisfy (i) and (ii) above) was defined as the minimum, over all ensembles $\{p_j, | \psi_j \rangle\}$ that decompose $\rho$, of $\sum_j p_j Q_j$ where $ Q_j \equiv \int_0^{2\pi} \!\! \langle \psi_j | (\Delta\hat{X}_{\mu})^2 | \psi_j \rangle \,d\mu/(2\pi)- 1/2$~\cite{Kwon19,YadinPRX18}. 
This resource measure, denoted by $\mathcal{Q}(\rho)$ in~\cite{Kwon19}, is the convex roof of the average quadrature variance of $\hat{\rho}$~\footnote{Here we define $\mathcal{Q}$ as $1/2$ of the value defined in~\cite{Kwon19}.}. While $\mathcal{Q}$ satisfies all four resource-measure properties above, it only quantifies the quantum-enhanced precision for pure states, for which it is the value of this precision averaged over all quadratures~\cite{Kwon19}. 

\textit{Operational resource measure of nonclassicality.---} Our operational resource measure is  
\begin{align}
\mathcal{N}\left(\hat{\rho}\right) & = \min_{\{p_j,\ket{\psi_j}\}}\biggl\{\max_{\mu}\sum_j p_j\langle \psi_j | (\Delta\hat{X}_{\mu})^2 | \psi_j \rangle \biggr\}-\frac{1}{2}\nonumber\\
&=\min_{\{p_j,\ket{\psi_j}\}} \biggl\{\sum_j p_j\left(\bar{n}_j-|\bar{\alpha}_j|^2\right)+\biggl|\sum_j p_j\left(\bar{\xi}_j-\bar{\alpha}_j^2\right)\biggr|\biggr\} ,  \label{eq:n-mixed}
\end{align}
in which the minimization is over all ensembles,  $\{p_j, \ket{\psi_j}\}$, that decompose $\hat{\rho}$. For convenience we have defined the moments $\bar{n}_j \equiv \bra{\psi_j}\hat{a}^{\dagger}\hat{a}\ket{\psi_j}$,  $\bar{\xi}_j \equiv \bra{\psi_j}\hat{a}^2\ket{\psi_j}$,  and $\bar{\alpha}_j \equiv \bra{\psi_j}\hat{a}\ket{\psi_j}$. With the help of a set of orthogonal vectors, $\{\ket{j}_E\}$, from an additional mode $E$, we can construct an extended state $\hat{\rho}_E\equiv\sum_j p_j \ket{\psi_j}\bra{\psi_j}\otimes\ket{j}_E\bra{j}$ allowing us to write $\mathcal{N}\left(\hat{\rho}\right)$ in the more compact form 
\begin{align}
\mathcal{N}\left(\hat{\rho}\right)=\min_{\{p_i,\ket{\psi_i}\}}\max_{\mu}F_{ \hat{X}_{\mu}\otimes \hat{I}_E}\left(\hat{\rho}_E\right)-\frac{1}{2}.\label{eq:extended}
\end{align}

We show that $\mathcal{N}(\hat{\rho})$ satisfies non-negativity and convexity using the expression in the second line in Eq.\eqref{eq:n-mixed}. Due to the special form in Eq.\eqref{eq:extended}, we can prove that $\mathcal{N}(\hat{\rho})$ satisfies weak monotonicity using the method provided in the supplement~\cite{SM}. 
Moreover, the measure $\mathcal{N}(\hat{\rho})$ also satisfies  
 \begin{align} 
      \mathcal{N}\left(\hat{\rho}\right) \ge \mathcal{W}(\hat{\rho}) . 
 \label{eq:ineq1} 
 \end{align}
For pure states $\mathcal{N}=\mathcal{W}$, which in this case can be written as
\begin{align}
    \mathcal{W}(|\psi\rangle) = \bar{n}-|\bar{\alpha}|^2+\left|\bar{\xi}-\bar{\alpha}^2\right| , 
    \label{Wpure}
\end{align} 
where we have defined $\bar{n} = \langle \hat{a}^{\dagger}\hat{a}\rangle$,  $\bar{\xi}=\langle \hat{a}^2\rangle$,  and $\bar{\alpha}=\langle \hat{a}\rangle$. The first part of $\mathcal{W}(|\psi\rangle)$, the expression  $\bar{n}-|\bar{\alpha}|^2$, describes the contribution of the energy that cannot be removed by displacement operations. It is also the averaged excess variance, which for pure states can be written as $\mathcal{Q}\left(\ket{\psi}\right)\equiv\langle\left(\Delta\hat{x}\right)^2\rangle/2+\langle\left(\Delta\hat{p}\right)^2\rangle/2-1/2$, in which $\hat{x} \equiv \hat{X}_{\pi/2}$ and $\hat{p} \equiv \hat{X}_0$~\cite{YadinPRX18, Kwon19, HilleryPRA89}. The second part of $\mathcal{W}(\ket{\psi})$, the expression $\left|\bar{\xi}-\bar{\alpha}^2\right|$, quantifies the squeezing of the state. 

Thus one of our main results is that we find the first operational resource theoretic measure $\mathcal{N}$ that both satisfies the essential requirements of a resource measure of nonclassicality, and has a strong relationship to the metrological power of quadrature variance.

 \begin{figure}[t]
\leavevmode\includegraphics[width = 0.95 \columnwidth]{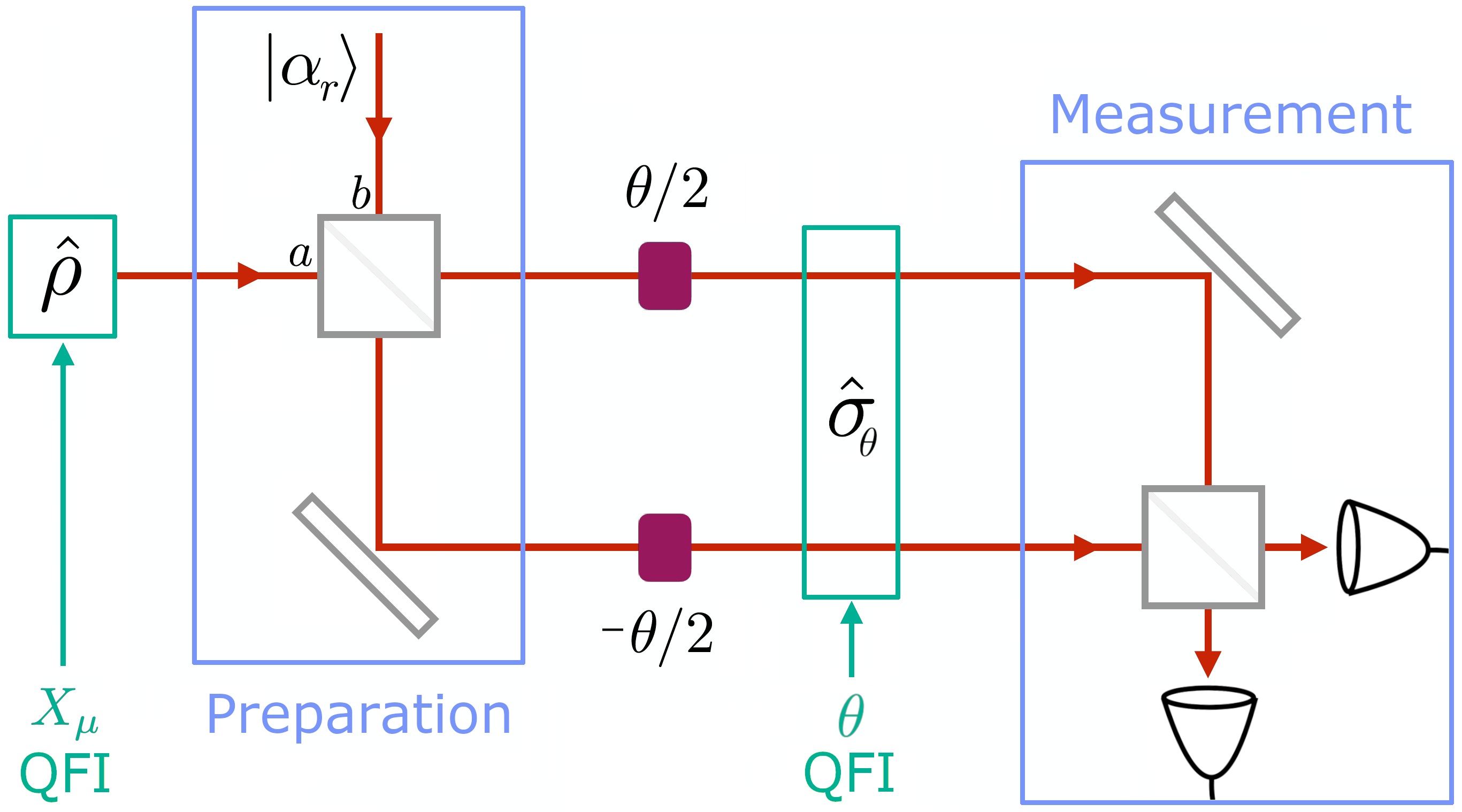}
\caption{Diagramatic depiction of a balanced Mach-Zehnder interferometer. In the part labelled ``preparation'' the input state $\hat\rho$ is combined with a coherent state $\ket{\alpha_r}$ at a 50/50 beamsplitter. The phase shift to be measured is then applied in an opposite direction to both output modes to create the two-mode state $\sigma_{\theta}$. Finally $\sigma_{\theta}$ is measured after combining both modes at a second 50/50 beamsplitter.} 
\label{fig:MZI} 
\end{figure} 

\emph{Precision of the Mach-Zehnder Interferometer.}---We now show that when we use the MZI of Fig.\ref{fig:MZI}, in which a coherent state $\ket{\alpha_r}$ is fed into one input and an arbitrary state $\hat\rho$ into the other, the optimal QFI, denoted by $F_{\theta}^{\text{MZI}}\left(\hat{\rho}\right)$, for determining the phase $\theta$ is directly related to that of $\hat\rho$ for quadrature sensing, $F_X(\hat\rho)$, defined for the metrological power $\mathcal{W}(\hat\rho)$. This fact has two immediate  consequences. First, it means that the MZI measures a phase shift by using the displacement sensitivity of the input state $\hat\rho$, which is quantified by $F_{X}(\hat\rho)$. Second, it means that if we replace the standard measurement configuration of the MZI with an optimal measurement, the MZI provides a measurement of the nonclassicality witness $\mathcal{W}(\hat\rho)$ and a lower bound on the nonclassicality $\mathcal{N}(\hat\rho)$. Note that one obtains a lower bound on $\mathcal{N}(\hat\rho)$ even with the standard MZI measurement. The optimal measurement merely provides a better lower bound.  

In an MZI, two input states are combined at a beam-splitter (BS), phase shifts $\theta_1$ and $\theta_2$ are applied respectively to the two outputs of the BS, and these outputs are then recombined on a second BS before being measured. This configuration provides a two-parameter metrology problem described by a 2x2 QFI matrix~\cite{Jarzyna12,Lang13}. It has been shown that when one input is a coherent state, the optimal precision in estimating the phase difference $\theta_1-\theta_2$ is provided by choosing the first BS to be 50/50 (in which case the MZI is referred to as balanced)~\cite{Jarzyna12,Hofmann:2009aa, SM}. By setting $\theta_1 = \theta/2 = -\theta_2$, which is the configuration shown in Fig.\ref{fig:MZI}, the optimal estimation can be equivalently described by the QFI for the single parameter $\theta$~\cite{Jarzyna12, SM}. For an unbiased estimator $\Theta$, the phase-estimation sensitivity satisfies the quantum Cramer-Rao bound \cite{Braunstein:1994aa}:
\begin{align}
\Delta^2\Theta\ge \frac{1}{MF_{\theta}^{\text{MZI}}\left(\hat{\rho}\right)},
\label{eq:CRB}
\end{align}
where $M$ is the number of repetitions. To show that the sensitivity to the phase $\theta$ for the MZI is related to that of the input $\hat\rho$ to a phase-space displacement, we use the explicit expression for the QFI~\cite{Braunstein:1994aa, toth2014quantum, Demkowicz-Dobrzanski:2015aa}. This involves an eigenstate decomposition of the density matrix, and is usually expressed in terms of a complete eigenbasis for this matrix. A useful expression in terms of those eigenstates with non-zero probabilities gives \cite{toth2014quantum}  
\begin{align}
    F_{\hat{G}}(\hat\sigma) = \mbox{Tr}[\hat{G}^2\hat\sigma] - \sum_{j,k} |\langle \phi_j | \hat{G} | \phi_k \rangle |^2 \left( \frac{2p_jp_k}{p_j+p_k}\right).
    \label{genQFI}
\end{align}
Here $\hat\sigma = \sum_{j=1}^M p_j |\phi_j\rangle\langle\phi_j |$, is the density matrix, $|\phi_j\rangle$ are the $M$ eigenstates of $\hat\sigma$ with non-zero eigenvalues $p_j$, and the sum over $j$ and $k$ runs from 1 to $n$. In our case, the operator that generates the phase shifts is $\hat{J}_z = \left(\hat{a}^\dagger \hat{a} - \hat{b}^\dagger \hat{b}\right)/2$, in which $\hat{a}$ and $\hat{b}$ are the modes for the input state $\hat\rho$ and the reference state $\ket{\alpha_r}$, respectively. It is simplest, however, to consider the initial BS as part of the applied transformation. In this picture, the state $\hat\sigma$ in Eq.(\ref{genQFI}) is the state input to the MZI, $\hat\sigma = \hat\rho \otimes |\alpha_r \rangle \langle \alpha_r|$, and the generator becomes $\hat{G} = \hat{U} \hat{J}_z \hat{U}^\dagger = (\hat{a} \hat{b}^\dagger - \hat{a}^\dagger \hat{b})/(2i)$, where $\hat{U} = \exp[-i\pi(\hat{a}^\dagger \hat{b} + \hat{a} \hat{b}^\dagger )/4]$ is the transformation applied by the BS. 

Returning to the expression for the QFI, Eq.(\ref{genQFI}), we note that the eigenstates of the input state $\hat\sigma$ are $|\phi_j\rangle = |\psi_j\rangle |\alpha_r\rangle$ in which the states $|\psi_j\rangle$ are the eigenstates of the input state $\hat\rho$. Taking the partial trace over the input mode that contains the coherent state $|\alpha_r\rangle$ gives
\begin{align}
    \mbox{Tr}[\hat{G}^2\hat\sigma] & = \frac{\mbox{Tr}[\hat{a}^\dagger \hat{a} \hat\rho]}{4} +  \frac{|\alpha_r|^2}{2} \mbox{Tr}[\hat{X}_{\phi}^2\hat\rho] , \\
    |\langle \phi_j | \hat{G} | \phi_k \rangle|^2 & = \frac{|\alpha_r|^2}{2} |\langle \psi_j | \hat{X}_{\phi} | \psi_k \rangle|^2 ,
\end{align}
in which $\alpha_r = |\alpha_r|e^{-i\phi}$. Substituting these into Eq.(\ref{genQFI}), and choosing $\phi$ to maximum the QFI \cite{Liu:2013aa}, we have 
\begin{align}
F_{\theta}^{\text{MZI}}\left(\hat{\rho}\right) & = \frac{\bar{n}}{4} + \frac{\left|\alpha_r\right|^2}{2} F_X\left(\hat{\rho}\right)= \frac{N}{4} +\frac{\left|\alpha_r\right|^2}{2}\left[F_X\left(\hat{\rho}\right)-\frac{1}{2}\right],  
\label{FF}
\end{align}
where $N \equiv \bar{n}+|\alpha_r|^2$ is the total mean number of photons input to the MZI, and $\bar{n}\equiv \mbox{Tr}[\hat{a}^\dagger \hat{a} \hat{\rho}]$ and $|\alpha_r|^2$ are the average numbers of photons in $\hat{\rho}$ and $\ket{\alpha_r}$, respectively. 

It follows from Eq.~\eqref{FF} that the MZI will achieve Heisenberg-limited phase sensing (that is, $F_{\theta}^{\text{MZI}}\left(\hat{\rho}\right)$ will scale as $N^2$~\cite{Pezze08}) if $F_X\left(\hat{\rho}\right)$ is proportional to $\bar{n}$ and we choose $|\alpha_r|^2 \sim \bar{n}$. This is true, for example, for all the classes of states for which $\mathcal{N}$ is shown in Table~\ref{tab}. These classes are: Fock states, $\ket{n}$ (the eigenstates of $a^\dagger a$ with eigenvalue $n$); a superposition of two Fock states $\ket{\phi(n)}\equiv \frac{1}{\sqrt{2}}\left(\ket{0}+\ket{n}\right)$ for $n>2$; squeezed vacuum states $\ket{\xi}\equiv \mathcal{S}(\xi)\ket{0}$ where $\mathcal{S}(\xi)=\exp\left(\frac{1}{2}\xi \hat{a}^{\dagger2}-\frac{1}{2}\xi^{\ast}\hat{a}^2\right)$; and ``cat" states $\ket{\alpha}_{\pm}= N_{\pm}^{-1/2}\left(\ket{\alpha}\pm\ket{-\alpha}\right)$ with $N_{\pm}=2\pm2 e^{-2|\alpha|^2}$. We note that for any pure state $|\psi\rangle$, using the Cauchy-Schwarz inequality one has $\left|\bar{\xi}\right|= \left|\braket{\psi|\hat{a}^2|\psi}\right|\le \sqrt{\braket{\psi|\hat{a}\hat{a}^{\dagger}|\psi}\braket{\psi|\hat{a}^{\dagger}\hat{a}|\psi}}=\sqrt{\bar{n}\left(\bar{n}+1\right)}$. One can also show that the maximum is achieved if and only if $|\psi\rangle$ is a squeezed state~\cite{SM}. Under the measure $\mathcal{N}$, the squeezed states are thus the most nonclassical and the most useful in the MZI for a fixed energy $\bar{n}$~\cite{Lang13}. 


By rearranging Eq.(\ref{FF}) we obtain the nonclassicality witness $\mathcal{W}$ directly in terms of the QFI for the MZI:
\begin{align}
\mathcal{W}\left(\hat{\rho}\right)  & = \max\left\{ \frac{F_{\theta}^{\ms{MZI}}\left(\hat{\rho}\right)-N/4}{|\alpha_r|^2/2},0\right\} . 
\label{eq:witness}
\end{align}
Since $\mathcal{W}$ is a necessary and sufficient condition for nonclassicality for all Gaussian states, the preparation part of the MZI provides a state, $\hat\sigma_{\theta}$ (see Fig. \ref{fig:MZI}), that is sufficient to fully determine the nonclassicality of that class~\cite{Aspachs09, Friis2015,Rig17}. However, to do so one must in general replace the measurement part of the MZI with an optimal measurement for the given input state $\hat{\rho}$. Explicit optimal measurement schemes have been determined for certain classes of Gaussian states~\cite{Seshadreesan:2011aa, Oh:2019aa}. For all pure states that have a real projection coefficient in the Dicke basis, the optimal measurement is achieved by recording the number of photons detected by each detector after the second BS~\cite{Lang13}. 

We stated above that the MZI provides a simple experimental procedure to obtain a lower bound on the nonclasicality, $\mathcal{N}(\rho)$. Combining Eqs. \eqref{eq:ineq1}, \eqref{eq:CRB}, and \eqref{eq:witness} provides the explicit expression for this lower bound in terms of the precision of the MZI, $\Delta^2\Theta$: 
\begin{align}
\mathcal{N}\left(\hat{\rho}\right)\ge \frac{4-NM\Delta^2\Theta}{2M|\alpha_r|^2\Delta^2\Theta}.
\end{align}

\renewcommand{\arraystretch}{1.4}
\setlength{\tabcolsep}{8pt}
 
\begin{table}
\caption{Nonclassicality for some classes of pure states. Definitions of these classes are given in the text.}
\begin{ruledtabular}
\begin{tabular}{c c c c c}
\label{tb}
& $\ket{n}$   & $\ket{\phi(n)}$  &  $\ket{\xi},\xi = re^{i\phi}$ & $\ket{\alpha}_{\pm}$ \\
\hline
$\mathcal{N}$ & $n$ & $n/2$ & $(e^r-1)/2$  & $|\alpha|^2\left(N_++N_-\right)/N_{\pm}$\\ 
$\mathcal{N}/\bar{n}$ & $1$  & $1$ & $1+\sqrt{1+1/\bar{n}}$ & $1+N_{\pm}/N_{\mp}$
\end{tabular}
\end{ruledtabular}
\label{tab}
\end{table}



\emph{Quantifying macroscopicity.}--- Creating superpositions of distinguishable states becomes harder as the ``size'' of the superposition increases. Here ``size'' may refer to the distance between the superposed states and/or the mass or energy of these states. For a linear oscillator,  and equivalently a single mode, the distance between the superposed states is closely related to the energy of these states, so we can expect good measures of the size, or \textit{macroscopicity}~\cite{Frowis2018} of a superposition to scale with energy. 

As discussed in the introduction, the notions of nonclassicality and macroscopicity are very similar, and it is therefore natural to examine how nonclassicality measures quantify the latter. Recent studies have considered the QFI~\cite{LeePRL11, Fr_wis_2012} and a measure of coherence~\cite{Yadin16} in this context. Here we show that $\mathcal{N}$ has an intuitive interpretation in terms of macroscopicity.


Consider a pure state $\ket{\psi}=\sum_{k=1}^L c_j\ket{\alpha_j}$ that is a superposition of the $L$ coherent states $\ket{\alpha_j}$. The complex amplitudes $\alpha_j$ and the coefficients $c_j$ satisfy the normalization condition $\sum_{j,k}c_jc_k^{\ast}\braket{\alpha_k|\alpha_j}=1$ (the amplitudes appear in this condition because the coherent states are not orthogonal). We present the full expression for  $\mathcal{N}$ in terms of $\alpha_j$ and $c_j$ in the supplement~\cite{SM}. As for any pure superposition of coherent states, the quantity $\bar{n}-|\bar{\alpha}|^2$ that appears in $\mathcal{N}$ (Eq.(\ref{Wpure})) is positive. Denoting the phase space distance between two coherent states $\ket{\alpha_j}$ and $\ket{\alpha_k}$ by $d_{jk}\equiv\alpha_j-\alpha_k$, and their overlap (inner product) by $f_{jk}\equiv\braket{\alpha_k|\alpha_j}$, we find that
\begin{align}
\bar{n}-|\bar{\alpha}|^2&=\frac{1}{2}\sum_{j,k}^L|c_j|^2|c_k|^2|d_{jk}|^2\left(1-|f_{jk}|^2\right)\nonumber\\
&+\sum_{j\ne k\ne l}^Lc_j^{\ast}|c_k|^2c_ld_{jk}^{\ast}d_{lk}\Big(f_{lj}-f_{lk}f_{kj}\Big)\nonumber\\
&+\frac{1}{2}\sum_{j\ne k\ne l\ne m}^Lc_j^{\ast}c_k^{\ast}c_lc_md_{jk}^{\ast}d_{ml}f_{mj}f_{lk}.
\label{eq:macro}
\end{align}
The expression for the other part of $\mathcal{N}$, namely $|\bar{\xi}-\bar{\alpha}^2|$, contains very similar terms~\cite{SM}.
The first line in Eq.(\ref{eq:macro}) contains contributions that come solely from pairs of coherent states, and the second and third lines from
triples and quadruples, respectively. When all the coherent states are far enough apart that they are approximately mutually orthogonal, all terms vanish except for the first term on
the first line. In that case $|\bar{\xi}-\bar{\alpha}^2|=\bar{n}-|\bar{\alpha}|^2$ and we have 
\begin{align}
\mathcal{N}\left(\ket{\psi}\right)=\sum_{j,k}^L|c_j|^2|c_k|^2|d_{jk}|^2.
\end{align}
Since we can interpret $P_{j,k} = |c_j|^2|c_k|^2$ as the probability of occurrence of each pair of coherent states, the non-classicality in this case is merely the mean square of the phase-space distances between pairs of coherent states. This gives $\mathcal{N}$ a very simple interpretation in phase space. 

 \emph{Summary.}--- Here we have presented the first operational resource-theoretic measure of nonclassicality, in which the operational component is the ability of a state to measure displacement in phase space. We have also shown that it provides a quantifier of macroscopicity with a simple geometrical interpretation in phase space.  Important open questions that will further the understanding of nonclasicality and its manipulation involve finding resource theoretic measures that quantify metrology of other quantities such as phase shifts, and as well as other technologically relevant tasks. 
 


\textit{Acknowledgments:} This research was supported by a grant from King Abdulaziz City for Science and Technology (KACST).
\\

\appendix
\quad \qquad \textbf{SUPPLEMENTAL MATERIALS}\\
In the supplemental materials, we provide detailed derivations about our results in the main manuscript. In Sec. \ref{sec1}, we prove the properties of our nonclassicality measure $\mathcal{N}$, including the requirements from the resource theory, some inequality relations, and the derivation of the most nonclassical state for a fixed energy. In Sec. \ref{sec2}, we prove that the \emph{balanced} MZI achieves the maximum quantum Fisher information (QFI) for an input state $\hat{\rho}\otimes\ket{\alpha_r}\bra{\alpha_r}$. In Sec. \ref{sec4}, we provide the full expression of nonclassicality in terms of coherent superpositions.

\section{Properties of our nonclassicality\label{sec1}}

\subsection{Resource theory requirements}
Here we prove several properties from the resource theory on our nonclassicality measure defined as
\begin{align}
\mathcal{N}\left(\hat{\rho}\right)&=\min_{\{p_j,\ket{\psi_j}\}}\left\{\max_{\mu}\sum_j p_j\left(\Delta\hat{X}_{\mu}\right)_{\psi_j}^2\right\}-\frac{1}{2}\nonumber\\
&=\min_{\{p_j,\ket{\psi_j}\}} \left\{\sum_j p_j\left(\bar{n}_j-|\bar{\alpha}_j|^2\right)+\left|\sum_j p_j\left(\bar{\xi}_j-\bar{\alpha}_j^2\right)\right|\right\}\nonumber\\
&=\min_{\{p_i,\ket{\psi_i}\}}\max_{\mu}F_{ \hat{X}_{\mu}\otimes \hat{I}_E}\left(\hat{\rho}_E\right)-\frac{1}{2},
\label{eq:n-mixed}
\end{align}
where $\hat{\rho}=\sum_j p_j \ket{\psi_j}\bra{\psi_j}$, $\hat{X}_{\mu}=i\left(e^{-i\mu}\hat{a}^{\dagger}-e^{i\mu}\hat{a}\right)/\sqrt{2}=\sin\mu \hat{x}+\cos\mu\hat{p}$, and  $\hat{\rho}_E\equiv\sum_j p_j \ket{\psi_j}\bra{\psi_j}\otimes\ket{j}_E\bra{j}$ with $\ket{j}_E$ orthogonal vectors in an additional mode $E$. The last line can be verified from the definition of the QFI $F_{ \hat{X}_{\mu}\otimes \hat{I}_E}\left(\hat{\rho}_E\right)$ of the quadrature $\hat{X}_{\mu}\otimes \hat{I}_E$ for the state 
$\hat{\rho}_E$. Same to the main text, we rescale the QFI by a factor of $4$.

\noindent \textit{(i) Non-negativity}: For an arbitrary state, it can be given by $\hat{\rho}=\sum_jp_j \ket{\psi_j}\bra{\psi_j}$ with $p_j>0$. For a classical state, $\mathcal{N}\left(\hat{\rho}\right)=0$ by decomposing the state using coherent states. For a nonclassical state, at least one of the decomposed state, say $\ket{\psi_s}$ , is not a coherent state by definition. Using Cauchy-Schwarz inequality, we have $|\bar{\alpha}_s|=|\bra{\psi_s}a\ket{\psi_s}|\le \sqrt{\bra{\psi_s}\ket{\psi_s}\bra{\psi_s}a^{\dagger}a\ket{\psi_s}}=\sqrt{\bar{n}_s}$, where the equality holds only if $a\ket{\psi_s}$ and $\ket{\psi_s}$ are parallel to each other, which means $\ket{\psi_s}$ is a coherent state \cite{SZ}. Since $\ket{\psi_s}$ is not a coherent state, we have $|\bar{\alpha}_s|<\sqrt{\bar{n}_s}$ and $\mathcal{N}\left(\hat{\rho}\right)>0$. Therefore, $\mathcal{N}\left(\hat{\rho}\right)>0$ if and only if when the state $\hat{\rho}$ is a nonclassical state. \\

\noindent \textit{(ii) Weak monotonicity}: An arbitrary linear optical mapping on the state $\hat{\rho}$ can be described by
\begin{align}
\Lambda\left(\hat{\rho}\right)&=\text{Tr}_A\left[\mathcal{U}\hat{\rho}\otimes \hat{\rho}_A \mathcal{U}^{\dagger}\right]=\sum_j\hat{K}_j\hat{\rho}\hat{K}_j^{\dagger}=\sum_jq_j\hat{\sigma}_j,
\end{align}
where $\hat{\rho}_A$ is a classical ancilla state, $\mathcal{U}$ is the linear optical unitary, and $\hat{K}_j$ are Kraus operators with $\sum_j \hat{K}_j^{\dagger}\hat{K}_j=\hat{I}$. The probabilities are $q_j=\text{Tr}\left(\hat{K}_j\hat{\rho}\hat{K}_j^{\dagger}\right)$, and the post-selected states are $\hat{\sigma}_j=\left(\hat{K}_j\hat{\rho}\hat{K}_j^{\dagger}\right)/q_j=\sum_{i}p_i\ket{\phi_{ij}}\bra{\phi_{ij}}$, where $\ket{\phi_{ij}}=\hat{K}_j\ket{\psi_j}/\sqrt{q_j}$ with a decomposition $\hat{\rho}=\sum_i p_i \ket{\psi_i}\bra{\psi_i}$. 

From our definition Eq. \eqref{eq:n-mixed}, we have the following inequality
\begin{align}
&\mathcal{N}\left[\Lambda\left(\hat{\rho}\right)\right]+\frac{1}{2}=\mathcal{N}\left(\sum_{i,j}p_iq_j\ket{\phi_{ij}}\bra{\phi_{ij}}\right)+\frac{1}{2}\nonumber\\
&\le \max_{\mu}F_{\hat{X}_{\mu}\otimes \hat{I}_E\otimes \hat{I}_{A^{\prime}}}\left(\sum_{i,j}p_iq_j\ket{\phi_{ij}}\bra{\phi_{ij}}\otimes\ket{i}_E\bra{i}\otimes\ket{j}_{A^{\prime}}\bra{j}\right),
 \label{eq:QFI-m} 
\end{align}
where $\ket{j}_{A^{\prime}}$ are orthogonal vectors on an ancilla mode $A^{\prime}$. We can choose the mode $E$ to be independent from the linear optical mapping $\mathcal{U}$ and the mode $A^{\prime}$. Since the mapping $\Lambda$ is linear and $\ket{i}_E$ are independent from the mapping, we can have $\Lambda\left(\hat{\rho}_E\right)=\sum_i p_i \Lambda\left(\ket{\psi_i}\bra{\psi_i}\right)\otimes\ket{i}_E\bra{i}$. According to Ref. \cite{Tan17}, there exists a classical ancilla state $\rho_{AA^{\prime}}$ and a linear optical unitary $\mathcal{U}$ such that $\text{Tr}_A\left[\mathcal{U}\hat{\rho}_E\otimes \hat{\rho}_{AA^{\prime}} \mathcal{U}^{\dagger}\right]=\sum_{i,j}p_iq_j\ket{\phi_{ij}}\bra{\phi_{ij}}\otimes\ket{i}_E\bra{i}\otimes\ket{j}_{A^{\prime}}\bra{j}$. Using the contractivity of partial trace on the QFI \cite{toth2014quantum}, the last line in Eq. \eqref{eq:QFI-m} can be upper bounded by 
\begin{align}
\max_{\mu}F_{\hat{X}_{\mu}\otimes \hat{I}_E\otimes\hat{I}{AA^{\prime}}}\left(\mathcal{U}\hat{\rho}_E\otimes \hat{\rho}_{AA^{\prime}} \mathcal{U}^{\dagger}\right).
\label{eq:monotone2}
\end{align}
By writing $\hat{\rho}_{AA^{\prime}}=\sum_kr_k \ket{\bm{\alpha}_k}\bra{\bm{\alpha}_k}$ with $\ket{\bm{\alpha}_k}$ a coherent state in the $AA^{\prime}$ modes and $\sum_kr_k=1$, we have 
$F_{\hat{X}_{\mu}\otimes \hat{I}_E\otimes\hat{I}{AA^{\prime}}}\left(\mathcal{U}\hat{\rho}_E\otimes \hat{\rho}_{AA^{\prime}} \mathcal{U}^{\dagger}\right)\le \sum_k r_kF_{\hat{X}_{\mu}\otimes \hat{I}_E\otimes\hat{I}{AA^{\prime}}}\left(\mathcal{U}\hat{\rho}_E\otimes \ket{\bm{\alpha}_k}\bra{\bm{\alpha}_k} \mathcal{U}^{\dagger}\right)$ from the convexity of the QFI. By transforming the operators, we have $\mathcal{U}\hat{X}_{\mu}\otimes \hat{I}_E\otimes\hat{I}{AA^{\prime}}\mathcal{U}^{\dagger}=\hat{X}^U_{\mu}+\hat{X}_{AA^{\prime}}$, where $\hat{X}^U_{\mu}=-ie^{i\mu}(U_{11}\hat{a}+h.c.)/\sqrt{2}$, $\hat{X}_{AA^{\prime}}=-ie^{i\mu}\sum^{N_A}_{j=2} U_{1j}\hat{a}_j/\sqrt{2}+h.c.$, and $\hat{a}_j$ are annihilation operators of the anncilla modes. According to the additivity of the QFI under tensoring \cite{toth2014quantum}, we obtain
\begin{align}
&\max_{\mu}F_{\hat{X}_{\mu}\otimes \hat{I}_E\otimes\hat{I}{AA^{\prime}}}\left(\mathcal{U}\hat{\rho}_E\otimes \ket{\bm{\alpha}_k}\bra{\bm{\alpha}_k} \mathcal{U}^{\dagger}\right)\nonumber\\
&= \max_{\mu}\left[F_{\hat{X}_{\mu}^U}\left(\hat{\rho}_E\right)+F_{ \hat{X}_{AA^{\prime}}}\left(\ket{\bm{\alpha}_k}\bra{\bm{\alpha}_k}\right)\right]\nonumber\\
&=|U_{11}|^2 \max_{\mu}F_{\hat{X}_{\mu}\otimes\hat{I}_E}\left(\hat{\rho}_E\right)+\frac{1}{2}\sum_{j=2}^{N_A}|U_{1j}|^2\nonumber\\
&=|U_{11}|^2 \max_{\mu}\left[F_{\hat{X}_{\mu}\otimes\hat{I}_E}\left(\hat{\rho}_E\right)-\frac{1}{2}\right]+\frac{1}{2}\nonumber\\
&\le \max_{\mu}F_{\hat{X}_{\mu}\otimes\hat{I}_E}\left(\hat{\rho}_E\right),
\label{eq:monotone3}
\end{align}
where we have used $|U_{11}|\le1$ and $\max_{\mu}F_{\hat{X}_{\mu}\otimes\hat{I}_E}\left(\hat{\rho}_E\right)\ge\frac{1}{2}$ in deriving the inequality. Therefore, $\mathcal{N}\left[\Lambda\left(\hat{\rho}\right)\right]\le \max_{\mu}F_{\hat{X}_{\mu}\otimes\hat{I}_E}\left(\hat{\rho}_E\right)-\frac{1}{2}$ with the state $\hat{\rho}_E=\sum_i p_i \ket{\psi_i}\bra{\psi_i}\otimes\ket{i}_E\bra{i}$ for any decomposition set $\{p_i,\ket{\psi_i}\}$. Choosing the minimum decomposition set, we obtain 
\begin{align}
\mathcal{N}\left[\Lambda\left(\hat{\rho}\right)\right]&\le \min_{\{p_i,\ket{\psi_i}\}}\max_{\mu}F_{\hat{X}_{\mu}\otimes\hat{I}_E}\left(\hat{\rho}_E\right)-\frac{1}{2}=\mathcal{N}\left(\hat{\rho}\right).
\label{eq:wm}
\end{align}
Hence we prove the weak monotonicity.

\noindent \textit{(iv) Convexity}: 
\begin{align}
&\sum_ip_i\mathcal{N}\left(\hat{\rho}_i\right)\nonumber\\
&=\sum_ip_i\left[\sum_j q^{\min}_{ij}\left(\bar{n}^{\min}_{ij}-|\bar{\alpha}^{\min}_{ij}|^2\right)+\left|\sum_j q^{\min}_{ij}\left(\bar{\xi}^{\min}_{ij}-(\bar{\alpha}^{\min}_{ij})^2\right)\right|\right]\nonumber\\
&\ge\sum_{i,j} p_iq^{\min}_{ij}\left(\bar{n}^{\min}_{ij}-|\bar{\alpha}^{\min}_{ij}|^2\right)+\left|\sum_{i,j} p_iq^{\min}_{ij}\left(\bar{\xi}^{\min}_{ij}-(\bar{\alpha}^{\min}_{ij})^2\right)\right|\nonumber\\
&\ge\mathcal{N}\left(\sum_ip_i\hat{\rho}_i\right),
\end{align}
where the equality is obtained from the definition of $\mathcal{N}$ with the superscript $\min$ representing the set of the minimum decomposition $\{q^{\min}_{ij},\ket{\psi^{\min}_{ij}}\}$ for each state $\hat{\rho}_i$. The last line is also due to the definition of $\mathcal{N}$ by recognizing that $\sum_{i,j}p_iq^{\min}_{ij}\ket{\psi^{\min}_{ij}}\bra{\psi^{\min}_{ij}}$ is one possible decomposition of the state $\sum_ip_i\hat{\rho}_i$.\\

\subsection{Inequalities with nonclassicality and the metrological power}
Recently, there is a nonclassicality quantifier $\mathcal{V}_1\left(\hat{\rho}\right)$ \cite{YadinPRX18} defined as the convex roof of the maximized quadrature variance, i.e., $\mathcal{V}_1\left(\hat{\rho}\right)=\min_{\{p_j,\ket{\psi_j}\}}\left\{\sum_j p_j\max_{\mu}\left(\Delta\hat{X}_{\mu}\right)_{\psi_j}^2\right\}-\frac{1}{2}$.
By definition, we find our measure
\begin{align}
\mathcal{N}\left(\hat{\rho}\right)=\min_{\{p_j,\ket{\psi_j}\}}\left\{\max_{\mu}\sum_j p_j\left(\Delta\hat{X}_{\mu}\right)_{\psi_j}^2\right\}-\frac{1}{2}\le\mathcal{V}_1\left(\hat{\rho}\right).
\end{align}

To prove the relation with the metrological power, we first show that $\mathcal{W}\left(\hat{\rho}\right)\le\mathcal{N}\left(\hat{\rho}\right)$  when $\mathcal{W}\left(\hat{\rho}\right)=0$. When $\mathcal{W}\left(\hat{\rho}\right)>0$, $\mathcal{W}\left(\hat{\rho}\right)
=\max_{\mu}F_{\hat{X}_{\mu}}\left(\hat{\rho}\right)-\frac{1}{2}$. We show that 
\begin{align}
\mathcal{N}\left(\hat{\rho}\right)&=\min_{\{p_i,\ket{\psi_i}\}}\max_{\mu}F_{\hat{X}_{\mu}\otimes\hat{I}_E}\left(\hat{\rho}_E\right)-\frac{1}{2}\nonumber\\
&\ge \min_{\{p_i,\ket{\psi_i}\}}\max_{\mu}F_{ \hat{X}_{\mu}}\left(\text{Tr}_E\left(\hat{\rho}_E\right)\right)-\frac{1}{2}\nonumber\\
&=\max_{\mu}F_{\hat{X}_{\mu}}\left(\hat{\rho}\right)-\frac{1}{2}=\mathcal{W}\left(\hat{\rho}\right).
\end{align}
Hence we prove the relation Eq. (6) in the main manuscript.

\subsection{A squeezed vacuum has the optimal metrological power}
We prove that the maximum $\mathcal{N}$ with the same mean number of photons $\bar{n}$ is achieved only by a squeezed vacuum $\ket{\xi}$, where $\xi=re^{i\phi}$. For any pure state,
\begin{align}
\mathcal{N}\left(\ket{\psi}\right)&=\bar{n}-|\bar{\alpha}|^2+\left|\bar{\xi}-\bar{\alpha}^2\right|\nonumber\\
&\le \bar{n}+\left|\bar{\xi}\right|= \bar{n}+\left|\braket{\psi|\hat{a}^2|\psi}\right|\nonumber\\
&\le \bar{n} +\sqrt{\braket{\psi|\hat{a}\hat{a}^{\dagger}|\psi}\braket{\psi|\hat{a}^{\dagger}\hat{a}|\psi}}\nonumber\\
&=\bar{n}+\sqrt{\bar{n}\left(\bar{n}+1\right)},
\end{align}
where the first inequality holds when $\alpha\equiv\braket{\psi|\hat{a}|\psi}=0$, and the second inequality holds when $\hat{a}\ket{\psi}$ is parallel to $\hat{a}^{\dagger}\ket{\psi}$, i.e., $\eta\hat{a}\ket{\psi}=\hat{a}^{\dagger}\ket{\psi}$ with $\eta$ a non-zero constant. By defining $\ket{\psi}=\sum_{n=0}^{\infty}c_n\ket{n}$, we obtain $\sum_{n=0}^{\infty}\eta c_n\sqrt{n}\ket{n-1}=\sum_{n=0}^{\infty} c_n\sqrt{n+1}\ket{n+1}$. This leads to a recursive relation 
\begin{align}
&c_{2n+1}=0, \quad \\
&c_{2n}=\eta\frac{\sqrt{2n-1}}{\sqrt{2n}}c_{2n-2},
\end{align}
where $c_0$ is determined through the normalization. We readily recognize the above coefficients as that of a squeezed vacuum with $\eta=e^{i\phi}\tanh r$ and $c_0=1/\sqrt{\cosh r}$. Moreover, $\bra{\xi}\hat{a}\ket{\xi}=0$. Therefore, only a squeezed vacuum achieves the maximum $\mathcal{N}$ for a fixed mean number of photons. For any mixed state $\hat{\rho}$ with $\bar{n}$, $\mathcal{N}\left(\hat{\rho}\right)<\bar{n}+\sqrt{\bar{n}\left(\bar{n}+1\right)}$ because it contains a mixture of pure states that does not achieve the maximum $\mathcal{N}$. Therefore, we prove that only a squeezed vacuum state $\ket{\xi}$ can achieve the maximum $\mathcal{N}=\bar{n}+\sqrt{\bar{n}\left(\bar{n}+1\right)}$ for a fixed $\bar{n}$.

\section{Phase estimation in the MZI \label{sec2}}
It is shown in Refs. \cite{Hofmann:2009aa, Jarzyna12} that a pure input state $\ket{\psi}\ket{\alpha_r}$ fed at a Mach-Zehnder interferometer (MZI) achieves the maximum quantum Fisher information when the interferometer is balanced, i.e., the transmission coefficient of the input beam-splitter $\tau=1/2$. This statement can be formulated as
\begin{align}
F_{\hat{J}_z}\left(\ket{\psi_{\theta,\tau}}\bra{\psi_{\theta,\tau}}\right)\le F_{\hat{J}_z}\left(\ket{\psi_{\theta,1/2}}\bra{\psi_{\theta,1/2}}\right),
\label{eq:max}
\end{align}
where $\ket{\psi_{\theta,\tau}}=\hat{U}_{\theta}\hat{U}_{\text{BS}}(\tau)\ket{\psi}\ket{\alpha_r}$ is the encoded state at the MZI, $\hat{U}_{\theta}=e^{-i\hat{J}_z\theta}$ with $\hat{J}_z = \left(\hat{a}^\dagger \hat{a} - \hat{b}^\dagger \hat{b}\right)/2$, and $\hat{U}_{\text{BS}}(\tau)=e^{-i\arcsin\sqrt{\tau}\left(a^{\dagger}b+b^{\dagger}a\right)}$. Here we show this statement is also true for a mixed input state of the form $\hat{\rho}\otimes \ket{\alpha_r}\bra{\alpha_r}$. For a mixed state $\hat{\rho}$, one can decompose it into
\begin{align}
\hat{\rho}=\sum_{j}p_j\ket{\psi_j}\bra{\psi_j},
\end{align}
where $p_j>0$ and $\sum_jp_j=1$. The encoded state is $\hat{\rho}_{\theta,\tau}=\hat{U}_{\theta}\hat{U}_{\text{BS}}(\tau)\hat{\rho}\otimes \ket{\alpha_r}\bra{\alpha_r}\hat{U}_{\text{BS}}^{\dagger}(\tau)\hat{U}_{\theta}=\sum_{j}p_j\ket{\psi_{\theta,\tau,j}}\bra{\psi_{\theta,\tau,j}}$, where $\ket{\psi_{\theta,\tau,j}}=\hat{U}_{\theta}\hat{U}_{\text{BS}}(\tau)\ket{\psi_j}\ket{\alpha_r}$. 

Therefore, the maximum QFI is at  $\tau=1/2$ since
\begin{align}
F_{\hat{J}_z}\left(\hat{\rho}_{\theta,1/2}\right)&= \min_{\{p_j,\ket{\psi_{\theta,j}}\}}\sum_j p_j F_{\hat{J}_z}\left(\ket{\psi_{\theta,1/2,j}}\bra{\psi_{\theta,1/2,j}}\right)\nonumber\\
&\ge \sum_j p^{\min}_j F_{\hat{J}_z}\left(\ket{\psi^{\min}_{\theta,\tau,j}}\bra{\psi^{\min}_{\theta,\tau,j}}\right)\nonumber\\
&\ge F_{\hat{J}_z}\left(\hat{\rho}_{\theta,\tau}\right),
\end{align}
where the equality is due to the convex roof construction of the QFI \cite{yu,TothPRA13}. The first inequality is obtained using Eq. \eqref{eq:max} with $\ket{\psi^{\min}_{\theta,\tau,j}}=\hat{U}_{\theta}\hat{U}_{\text{BS}}(\tau)\ket{\psi^{\min}_j}\ket{\alpha_r}$ and ${\{p^{\min}_j,\ket{\psi^{\min}_{\theta,j}}\}}$ denoting the convex roof decomposition, and the second inequality is due to the convexity of the QFI \cite{toth2014quantum}. Hence we have proved that with an arbitrary input state $\hat{\rho}\otimes \ket{\alpha_r}\bra{\alpha_r}$, the balanced MZI gives the maximum QFI.

\section{Relation to Quantum Macroscopicity\label{sec4}}
Here we provide the full expression of $\mathcal{N}$ for a pure state $\ket{\psi}=\sum_{j=1}^L c_j\ket{\alpha_j}$ in terms of the phase-space distance $d_{jk}\equiv\alpha_j-\alpha_k$ and the inner product $f_{jk}=\braket{\alpha_k|\alpha_j}$. We find $\mathcal{N}=\bar{n}-|\bar{\alpha}|^2+|\bar{\xi}-\bar{\alpha}^2|$, where 
\begin{align}
\bar{n}-|\bar{\alpha}|^2&=\frac{1}{2}\sum_{j,k}^L|c_j|^2|c_k|^2|d_{jk}|^2\left(1-|f_{jk}|^2\right)\nonumber\\
&+\sum_{j\ne k\ne l}^Lc_j^{\ast}|c_k|^2c_ld_{jk}^{\ast}d_{lk}\Big(f_{lj}-f_{lk}f_{kj}\Big)\nonumber\\
&+\frac{1}{2}\sum_{j\ne k\ne l\ne m}^Lc_j^{\ast}c_k^{\ast}c_lc_md_{jk}^{\ast}d_{ml}f_{mj}f_{lk},
\end{align}
and the squeezing term as
\begin{align}
|\bar{\xi}-\bar{\alpha}^2|&=\frac{1}{2}\left|\sum_{j,k}^L|c_j|^2|c_k|^2d_{jk}^2\left(1+2\frac{c_k^{\ast}}{c_j^{\ast}}f_{kj}+|f_{kj}|^2\right)\right|\nonumber\\
&+\sum_{j\ne k\ne l}^Lc_j^{\ast}|c_k|^2c_ld^2_{lk}\Big(f_{lj}+f_{lk}f_{kj}+\frac{c_j^{\ast}}{2c_k^{\ast}}f_{kj}f_{lj}\Big)\nonumber\\
&+\frac{1}{2}\sum_{j\ne k\ne l\ne m}^Lc_j^{\ast}c_k^{\ast}c_lc_md_{ml}^{2}f_{mj}f_{lk}
\label{eq:sq}
\end{align}
From the above equations, we see that the squeezing term $|\bar{\xi}-\bar{\alpha}^2|$ contains three contributions similar to that of $\bar{n}-|\bar{\alpha}|^2$. The first line on the right-hand side of Eq. \eqref{eq:sq} corresponds to the contribution from any two components in the superposition state, while the next two lines represent the contributions from any three and any four components, respectively. We note that for any coherent superposition state $\bar{n}-|\bar{\alpha}|^2$ is always non-zero, while it may not be true for $|\bar{\xi}-\bar{\alpha}^2|$. For example, the odd cat state $\ket{\alpha}_{-}=1/\sqrt{N_{-}}\left(\ket{\alpha}-\ket{-\alpha}\right)$ with $N_{-}=2-2 e^{-2|\alpha|^2}$ when $\alpha\rightarrow 0$.

When the coherent components are far apart from each other, they are almost orthogonal to each other such that $f_{jk}\approx0$. In this case, we have $\bar{n}-|\bar{\alpha}|^2=|\bar{\xi}-\bar{\alpha}^2|$ and the nonclassicality measure is
\begin{align}
\mathcal{N}\left(\ket{\psi}\right)=\sum_{j,k}^L|c_j|^2|c_k|^2|d_{jk}|^2.
\end{align}

\bibliography{Nonclassical-Measure}

\end{document}